\documentclass[aps,prb,preprint,showpacs]{revtex4}
\usepackage{graphicx}

\begin{document}
\def\he4{$^{4}$He}
\def\beq{\begin{equation}}
\def\eeq{\end{equation}}
\def\beqn{\begin{eqnarray}}
\def\eeqn{\end{eqnarray}}
\markboth{Massimo Boninsegni}
{Permutation Sampling in Path Integral Monte Carlo}
\title{Permutation sampling in Path Integral Monte Carlo}
\author{Massimo Boninsegni}

\affiliation{Department of Physics, University of Alberta\\
Edmonton, Alberta T6G 2J1,
Canada}

\date{\today}
\begin{abstract}
A simple algorithm is described to sample  permutations of identical particles in Path Integral  Monte Carlo (PIMC) simulations of  continuum many-body systems. The sampling strategy illustrated here is fairly general, and can be easily incorporated in any PIMC implementation based on the staging algorithm. Although it is similar in spirit to an existing prescription, it  differs from it in some key aspects. It allows one to sample permutations efficiently, even if long paths (e.g., hundreds, or thousands of slices) are needed.  We illustrate its effectiveness by presenting results of a PIMC calculation of thermodynamic properties of  superfluid  $^4$He, in which a very simple approximation for the high-temperature density matrix was utilized. 
\end{abstract}
\maketitle

\section{Introduction}
\label{intro}

The Path Integral Monte Carlo method is arguably the most powerful numerical technique to calculate thermodynamic properties of continuum (i.e., non-lattice) quantum many-body systems at finite temperature.\cite{ceperley95} For Bose systems, it is the only known, generally applicable theoretical method essentially free from approximations. Numerical estimates yielded by PIMC are affected by a statistical error, as well as by systematic errors, due to the finite size of the simulated system and to imaginary time discretization. In most cases, however, the computational resources typically available nowadays allow one to render the size of  all of these errors insignificant in practice.

The most notable application of PIMC to date, is the study of the superfluid (SF) transition in liquid $^4$He by Ceperley and Pollock,\cite{ceperley86} whose results have become the standard reference for {\it all} theoretical calculations on SF helium;  but numerous applications to other Bose systems have been reported in the literature, over the past two decades. \hfil\break No general formulation exists as yet of PIMC (nor of any other Quantum Monte Carlo method),   capable of overcoming the  {\it sign} problem, that has so far made it impossible to obtain equally high quality results for Fermi systems. Even for fermions, however, PIMC proves a valid option, allowing one to obtain approximate estimates, of accuracy at least comparable to that afforded by other methods.\cite{ceperley96}

Physical effects of interest in quantum many-body systems are almost invariably associated with quantum statistics; for example, superfluidity in $^4$He is intimately connected to long exchange cycles of helium atoms. Because a direct summation of all $N!$ permutations of $N$ indistinguishable particles is unfeasible, except for very small values of $N$, within PIMC  quantum statistics is included by performing  a statistical sampling of permutations. Thus, an all-important ingredient of any practical implementation of PIMC is an efficient procedure to carry out such a sampling.  

Since the pioneering study of Ref. \onlinecite{ceperley86}, there has been relatively little experimentation with implementations of PIMC differing, in some of the more important aspects, from the one described in Ref. \onlinecite{ceperley95}, henceforth referred to as CP. The CP implementation has come to be regarded as ``canonical", especially when studying quantum many-body systems in the highly degenerate regime (i.e. at low temperature). It is based on an accurate (``pair-product") high-temperature density matrix, allowing one to observe convergence of the physical estimates with a relatively low number of imaginary time ``slices" (of the order of 40 for superfluid $^4$He at a temperature $T$=1 K). Two slightly different procedures have been proposed and utilized, in the context of CP, to perform the sampling of permutations, both of which are thoroughly described in Ref. \onlinecite{ceperley95}. To our knowledge, no systematic, quantitative assessment of the relative merits and advantages of these two sampling strategies has yet been offered; it is also unclear to what extent their effectiveness and applicability are problem-dependent, and/or hinge on the use of the above-mentioned high-temperature density matrix.

In this work, we illustrate a new method to sample permutations of indistinguishable particles in PIMC simulations. It bears some similarities with one of the two strategies described in Ref. \onlinecite{ceperley95}, but differs from it in some important technical aspects. We also deem it easier to implement, and may be potentially more efficient, even though, naturally, this speculation will need to be supported by systematic comparisons with the other existing options.

As an illustrative application of our sampling method, we have carried out a PIMC simulation of  liquid $^4$He in the SF regime, i.e., we have repeated the original calculation of Ref. \onlinecite{ceperley86}. SF helium is the accepted test bench for quantum Monte Carlo calculations, since it is the most extensively studied quantum fluid, for which effects of quantum statistics manifest themselves at the macroscopic level.
In order to make the test more significant, and help expose any deficiency or merit of the permutation sampling procedure, we have {\it not} utilized  the pair-product high-temperature density matrix; rather we opted for a much simpler form, which requires a substantially larger number of imaginary time slices, in order to observe convergence of the estimates. Besides providing results for energetic properties, known to be affected quantitatively by Bose statistics, and which we compare to those of Ref. \onlinecite{ceperley86}, we have also attempted to  furnish here some quantitative information, which should help assess the efficiency of our method in sampling the space of all possible entangled many-particle paths (i.e., including permutations). It is our hope that this will provide a baseline for future, more extensive comparisons of different approaches. Somewhat interestingly, our results show that the PIMC simulation of SF  \he4 is feasible, albeit at a higher computational cost, even with a relatively simple PIMC implementation. Doubtless, this is also in part due to advances in computing hardware, which enable what may have been prohibitive two decades ago, when the first such simulation was carried out.

The remainder of this manuscript is organized as follows: in the next section, we provide a detailed description of our computational methodology.   In the following sections,  we illustrate our results, and outline our conclusions and outlook.

\section{METHODOLOGY}\label{meth}
The PIMC methodology is fairly mature, and extensively described in Ref. \onlinecite{ceperley95}, to which we refer readers interested in a thorough, comprehensive illustration. Our specific implementation is largely based on the ideas and methods presented therein. Nevertheless, at the risk of some redundancy, we provide a somewhat detailed description of our implementation here. This will hopefully facilitate the task of others who may wish to repeat our study and/or experiment with our algorithm to sample permutations.  

Consider a quantum many-body system of $N$ identical, point-like particles of mass $m$, described by the following Hamiltonian:
\beq\label{ham}
\hat H = -\lambda \sum_{i=1}^{N}\nabla^{2}_{i} +  \sum_{i<j} V(|{\bf r}_{i}-{\bf r}_{j}|)
\eeq
where $\lambda=\hbar^2/2m$. 
Implicit in the above model is the assumption that the interactions among particles can be accurately represented by a pairwise, central potential (the $V$ term in (\ref{ham})). Although this is obviously an approximation, it is commonly made in theoretical studies of most quantum fluids. In any case, it is not a requirement for the applicability of PIMC, nor of our specific implementation.
In the following, it is assumed that particles in the system obey Bose statistics.\cite{noteq} The system is assumed to be enclosed in a vessel, shaped as a parallelepiped, with periodic boundary conditions in all directions, and to be held in thermal equilibrium at a temperature $T$.

The thermodynamic average of a physical quantity formally represented by an operator $\hat {O}$ (for simplicity assumed diagonal in the coordinate representation)  is expressed as follows:
\beqn\label{ave}
\langle {O} \rangle = 
{1 \over Z} \ \int \ d{R} \ {O} ({R}) \ 
\rho({R},{R},\beta) 
\eeqn
where $\beta$=1/$T$ (we work with units where the Boltzmann constant $k_{B}$=1), and ${R} 
\equiv \{ {\bf r}_1,{\bf r}_2,...{\bf r}_N \} $,  is a configuration of the system, specified by the 
positions of all the $N$ particles. In Eq. (\ref{ave}), $\rho({R}, {R^\prime},\beta) \equiv \langle 
{R}| e^{-{\beta}{\hat H}} |{R^\prime} \rangle$ is  the many-body density 
matrix, and $Z = \int \ d{R} \ \rho({R}, {R},\beta)$ is the 
partition function.

A Monte Carlo evaluation of (\ref{ave}) consists of generating a large set of random many-particle configurations $\{R_p \}$, $p=1,...,M$, statistically sampled from a probability  density  proportional to $\rho({R},{R},\beta)$; the thermal average (\ref{ave}) can thus be estimated as a statistical average over the set of values 
$\{ {O}({R}_p) \}$. \

An explicit expression for the density matrix $\rho(R,R,\beta)$ is unavailable for any non-trivial many-body system; however, one can still generate the set $\{R_p\}$, by sampling {\it discrete many-particle paths} $X_p$ through configuration space, i.e.,
 \beq X_p\equiv \{R_{0p}, R_{1p}, R_{2p}, ...\ R_{Lp}\}\eeq 
 
Paths are formally defined in the {\it imaginary time} interval $0\le\tau\le\beta$, i.e., $R_{j}\equiv R(j\delta\tau)$, with $L\delta\tau=\beta$, and are randomly drawn from a probability distribution $\bar\rho(X)$ given by
\begin{equation}\label{paths}
\bar\rho(X)\equiv \bar \rho(R_{0}, R_{1}, R_{2}, ... R_{L}) = \prod_{j=0}^{L-1} \rho_\circ(R_{j},R_{j+1},\delta\tau)
\end{equation}
where $\rho_\circ$ is an (analytically known) approximation to the true many-body density matrix, constructed to be asymptotically exact in the ``high temperature'' $\delta\tau\to 0$ limit.  It can be shown that in that limit ($L\to\infty$), each configuration $R_{p}$ visited by  paths is statistically sampled from a distribution proportional to $\rho(R,R,\beta)$. 

The configuration $R(\beta)$, i.e., that corresponding to the end of the path in imaginary time, must coincide with $R(0)$, {\it except} for a permutation $P$ of the particle labels ($1$ through $N$). The possibility of permutations of particles must be allowed, in order to incorporate in the calculation the effects of particle indistinguishability and Bose statistics. Consequently, although many-particle paths are periodic in imaginary time, i.e., the configuration $R(j\delta\tau+q\beta)$ ($q$ being an arbitrary integer) is identical with $R(j\delta\tau)$ (in that particles occupy identical positions), individual particle labels can be different. Stated differently, single particle paths ${\bf r}_{1}(j\delta\tau)...{\bf r}_{N}(j\delta\tau)$ can become ``entangled'', as a result of permutations.

Permutations normally become important at sufficiently low temperature; at high temperature, only the identity permutation contributes significantly to thermal averages. At low temperature, however, permutations underlie phenomena such as superfluidity and Bose Condensation in liquid helium and, presumably, in all other superfluids. 

In an actual calculation implementing the above computational scheme, one must necessarily work with a finite value of $L$; in principle, one ought to carry out the $L\to \infty$ limit by extrapolating numerical results obtained with different values of $L$. In practice, this procedure proves quite cumbersome, especially when one is interested in many thermodynamic points. Thus, one typically performs all calculations (at a given temperature) with a single value of $L$, chosen sufficiently large so that estimates may be expected to coincide with the extrapolated values, within some small tolerance. 
For reasons of efficiency, it is desirable that such ``optimal" value of $L$ not be too large (a few hundred slices at the most); thus, it is advantageous to work with as accurate as possible  a ``high-temperature density matrix" $\rho_\circ$, which will allow one to achieve convergence of the numerical estimates without resorting to impractically large values of $L$. \\ The importance of this issue was demonstrated by Ceperley and Pollock,\cite{ceperley86} who proposed the following form for $\rho_\circ$:
\beq\label{ppa}
\rho_\circ(R,R^\prime,\delta\tau)=A_F(R,R^\prime,\delta\tau)\ \biggl \{ \prod_{i < j}
{\rm exp}\biggl [-u({\bf r}_{ij},{\bf r}^\prime_{ij}, \delta\tau)\biggr ]\biggr \}
\eeq
where ${\bf r}_{ij}={\bf r}_j-{\bf r}_i$, $A_F(R,R^\prime,\delta\tau)= \prod_{i=1}^N \rho_F({\bf r}_i,{\bf r}^\prime_i,\delta\tau)$, is the exact density matrix of a system of $N$ {\it distinguishable}, {\it non-interacting} particles, with
\beq
\rho_{F}({\bf r}_i,{\bf r}^\prime_i,\delta\tau)=
\biggl ( \sqrt {4\pi\lambda\delta\tau}\biggr )^{-3/2} \ {\rm exp}
\biggl [ -\frac {({\bf r}_i-{\bf r}^\prime_i)^2}{4\lambda\delta\tau}\biggr ]
\eeq
and where $u$ is obtained by imposing that $\rho_\circ$ be the exact density matrix for a system of two interacting particles. For PIMC calculations of highly quantal, hard-sphere-like systems such as condensed Helium, the form (\ref{ppa}) for $\rho_\circ$ affords a tremendous increase in efficiency, with respect  to other, simpler forms for $\rho_\circ$ (such as the so-called {\it primitive} approximation; for details, see Ref. \onlinecite{ceperley95}). 

In this work, we have {\it not} made use of the high-temperature density matrix (\ref{ppa}), choosing instead the following form:
\beq\label{npa}
\rho_\circ(R_j,R_{j+1},\delta\tau)= A_F(R_j,R_{j+1},\delta\tau)
  \ {\rm exp} \biggl [-\delta\tau U(R_{j})\biggr ]
\eeq
where
\beq \label{voth}
U(R_{j}) =
\frac{2V(R_{j})} {3}+{\tilde V}(R_{j})
\eeq
$V(R)\equiv\sum_{i<j} V(|{\bf r}_{i}-{\bf r}_{j}|)$ being the total potential energy of the system in the configuration $R_{j}$, and 
\begin{eqnarray} {\tilde V}(R_{j}) & = &
\frac{2{V}(R_{j})}{3}+\frac{2\lambda(\delta\tau)^2} {9}\sum_{i=1}^N
(\nabla_{i}{V}(R_{j}))^2 \end{eqnarray}
if $j$ is odd, and zero if $j$ is even. Here, $\nabla_{i}V(R)$ is the gradient of the total potential energy for the configuration $R$, with respect to the coordinates of the $i$th particle. 
This is a particular case of a more general expression, which can be shown\cite{moron,voth} to be accurate up to terms of order $\tau^4$ in the expansion of the exact density matrix $\rho(R,R^\prime,\tau)$ in powers of $\tau$.  

Using the form (\ref{voth}) instead of  the (superior) pair-product approximation (\ref{ppa}), results in a substantially larger value of $L$ required to achieve convergence. For the specific physical system that we have chosen to test our algorithm, namely superfluid \he4, the number $L$ of imaginary time slices needed is as much as 16 times greater than if (\ref{ppa}) had been used.
The reason for our choice is that  our interest in primarily methodological. Specifically, we wish to separate the relative contributions to the effectiveness of a PIMC implementation, of the permutation sampling procedure and of the high-temperature density matrix utilized.  A more stringent test is provided of our sampling scheme, if it can be shown to work satisfactorily with a fairly simple approximation for $\rho_{\circ}$.

\subsection{Path Sampling}
The generation of the set of many-particle paths $\{X_p\}$, with $p=1,2,... M $, can be conveniently carried out using the Metropolis algorithm. According to the standard procedure,\cite{allen} one performs a random walk through the space of $N$-particle paths $X$, defined above, starting from an initial  point $X_\circ$. The $X_p$'s are then the points sequentially visited by the random walk.

Let  $X_l$ be the $l$th element of the set $\{X_p\}$; in order to generate $X_{l+1}$, one samples a modification of the path $X_l$, involving  new positions of one or more particles at several  points (i.e., configurations) along $X_l$. Let $X_l^{\star}$ be the path  arising from such modification of $X_l$, and let $T(X_l\to X_l^{\star})$ be the probability with which $X_l^{\star}$ is sampled from $X_l$. The proposed new path is accepted, thereby becoming the next point of the random walk (as well as the next element $X_{l+1}$ of the set $\{X_p\}$), with probability
\beq\label{basic}
W(X_l \to X_l^{\star}) \equiv \frac {\bar \rho(X_l^\star)}{\bar \rho(X_l)}\ 
\frac{T(X_l^{\star}\to X_l)}{T(X_l\to X_l^{\star})}
\eeq
This is simply done by drawing a random number $\chi$  between zero and one;  if $W(X_l\to X_l^{\star}) > \chi$, then $X_{l+1} \equiv X_l^{\star}$, otherwise $X_{l+1}\equiv X_l$.

Of fundamental importance to the efficiency, unbiasedness and correctness  of the algorithm, are the elementary {\it moves} whereby one generates the ``trial" path $X_l^{\star}$ starting from $X_l$. In our PIMC implementation,  two different types of moves are performed. A detailed descriptions of these moves is offered in the next two subsections.
\subsubsection{``Wiggle" type moves}
These moves modify the current path $X_{l}$ by just altering the path of one particle, randomly chosen. Random displacements are applied  to a number $s-1$ of consecutive positions of that particle along its path. This can be thought of as ``chopping off" a portion of path, and replacing it with a different segment.  The maximum number of positions modified by the update is $L-1$, as the two ends of the portion are left unchanged.  Because paths are periodic, it is possible to update  a portion of path of a single particle that will include the  zeroth or $L$th positions.\cite{noteb}

In order to illustrate this type of move in detail, let us assume that a particle has been selected, and let the portion of path to be updated include the positions ${\bf r}_{k+1}...{\bf r}_{k+s-1}$, where $0\le k< L-1$ is an integer number, and $s=2^{m}$ is chosen so that $2 \le s\le L$. 
Let ${\bf r}^{\prime}_{k+1}...{\bf r}^{\prime}_{k+s-1}$ be the tentative new positions of the particle, selected according to some (yet unspecified) probabilistic criterion, expressed by a sampling function $T$. For notation purposes, we also define ${\bf r}^{\prime}_{k}={\bf r}_{k}$ and ${\bf r}^{\prime}_{k+s}={\bf r}_{k+s}$.
Based on (\ref {voth}) and (\ref{basic}), the acceptance probability of the move will be
\beq\label{wig}
W = \frac{\biggl \{ \prod_{j=0}^{s-1}\rho_{F}({\bf r}^{\prime}_{k+j},{\bf r}^{\prime}_{k+j+1},\delta\tau)\biggr \} \ {\rm exp}\biggl [ {-\delta\tau \sum_{j=1}^{s-1}U(R^{\prime}_{k+j})} \biggr ]}
{\biggl \{ \prod_{j=0}^{s-1}\ \rho_{F}({\bf r}_{k+j},{\bf r}_{k+j+1},\delta\tau)\biggr \}\  {\rm exp}\biggl [
-\delta\tau \sum_{j=1}^{s-1} U(R_{k+j})\biggr ] } \ 
\frac{T(X^{\prime}\to X)}{T(X\to X^{\prime})}
\eeq
having defined $R^{\prime}$ as the configuration that differs from $R$ only by the displacement of the chosen particle from ${\bf r}$ to ${\bf r}^{\prime}$, $X\equiv R_{0},R_{1},...R_{L}$ is the current path, whereas 
$X^{\prime}\equiv R_{0},R_{1},...R_{k},R^{\prime}_{k+1},...R^{\prime}_{k+s-1},R_{k+s},...R_{L}$ is the proposed new path.
There is considerable freedom in choosing the sampling probability $T$,\cite{allen} but it is clearly advantageous to do so in a way that will simplify the expression (\ref{wig}). An obvious choice is
\beqn\label{choose}
T(X\to X^{\prime}) = \prod_{j=0}^{s-1}\rho_{F}({\bf r}^{\prime}_{k+j},{\bf r}^{\prime}_{k+j+1},\delta\tau)
\eeqn
which reduces the acceptance probability (\ref{wig})  to 
\beq\label{wigg} W={\rm exp}\biggl [-\delta\tau \sum_{j=1}^{s-1}\biggl (U(R^{\prime}_{k+j})-U(R_{k+j})\biggr )\biggr]\eeq 
The probability $T$ so defined can be conveniently sampled by means of the ``staging'' algorithm.\cite{pollock84} The idea is as follows: $T$ is given by  a product of $s=2^{m}$ Gaussian terms, namely
\beqn\nonumber
T(X\to X^{\prime}) \propto {\rm exp}\biggl [{-\frac{({\bf r}^{\prime}_{k}-{\bf r}^{\prime}_{k+1})^{2}}{4\lambda\delta\tau}}\biggr ]\ \times \ 
{\rm exp}\biggl [{-\frac{({\bf r}^{\prime}_{k+1}-{\bf r}^{\prime}_{k+2})^{2}}{4\lambda\delta\tau}}\biggr ]\ ...\\
... \times\ {\rm exp}\biggl [{-\frac{({\bf r}^{\prime}_{k+s-1}-{\bf r}^{\prime}_{k+s})^{2}}{4\lambda\delta\tau}}  \biggr ]
\eeqn

Using some simple algebra\cite{note2} it is possible to re-organize this product in the following, ``hyerarchical'' form:
\begin{eqnarray}\label{casino}\nonumber
T \propto 
{\rm exp}\biggl [{-\frac{({\bf r}^{\prime}_{k}-{\bf {r}}^{\prime}_{k+s})^{2}}{4s\lambda\delta\tau}}\biggr ] \ \times\ 
{\rm exp}\biggl [{-\frac{({\bf r}^{\prime}_{k+s/2}-{\bf {\bar r}}^{\prime}_{k,k+s})^{2}}{s\lambda\delta\tau}} \biggr ] 
\\ \nonumber 
\times \  \biggl \{ {\rm exp}\biggl [{-\frac{({\bf r}^{\prime}_{k+s/4}-{\bf {\bar r}}^{\prime}_{k,k+s/2})^{2}}{s\lambda\delta\tau/2}}\biggr ]\ 
{\rm exp}\biggl [{-\frac{({\bf r}^{\prime}_{k+3s/4}-{\bf {\bar r}}^{\prime}_{k+s/2,k+s})^{2}}{s\lambda\delta\tau/2}}\biggr ] \biggr \} \\ \nonumber \times \ \biggl \{
{\rm exp}\biggr [{-\frac{({\bf r}^{\prime}_{k+s/8}-{\bf {\bar r}}^{\prime}_{k,k+s/4})^{2}}{s\lambda\delta\tau/4}}\biggr ]\
{\rm exp}\biggl [{-\frac{({\bf r}^{\prime}_{k+3s/8}-{\bf {\bar r}}^{\prime}_{k+s/4,k+s/2})^{2}}{s\lambda\delta\tau/4}}\biggr ]  \\ \nonumber   
{\rm exp}\biggl [{-\frac{({\bf r}^{\prime}_{k+5s/8}-{\bf {\bar r}}^{\prime}_{k+s/2,k+3s/4})^{2}}{s\lambda\delta\tau/4}} \biggr ]
\ {\rm exp}\biggl [{-\frac{({\bf r}^{\prime}_{k+7s/8}-{\bf {\bar r}}^{\prime}_{k+3s/4,k+s})^{2}}{s\lambda\delta\tau/4}} \biggr ]\biggr \} \\ \times\  \biggl \{
{\rm exp}\biggl [{-\frac{({\bf r}^{\prime}_{k+s/16}-{\bf {\bar r}}^{\prime}_{k,k+s/8})^{2}}{s\lambda\delta\tau/8}}\biggr ]\ ...\ {\rm etc}\ ... \biggr \}
\end{eqnarray}
where we have defined ${\bf {\bar r}}^{\prime}_{\alpha,\beta}\equiv ({\bf r}^{\prime}_{\alpha}+{\bf r}^{\prime}_{\beta})/2$. All distances are assumed to be computed with periodic boundary conditions, using the minimum image convention.
Expression (\ref{casino}) immediately suggests a sequential, {\it multi-level}  procedure to generate trial random positions of the particle being displaced. Since ${\bf r}^{\prime}_{k}={\bf r}_{k}$ and ${\bf r}^{\prime}_{k+s}={\bf r}_{k+s}$, the first factor in (\ref {casino}) does not enter the sampling in this type of move. Thus, one starts by generating a new ``midpoint'' position ${\bf r}^{\prime}_{k+s/2}$, by sampling a three-dimensional Gaussian distribution function of semi-width $\sigma_{0}=\sqrt{s\lambda\delta\tau/2}$, centered at 
$({\bf r}^{\prime}_{k}+{\bf r}^{\prime}_{k+s})/2$. It is customary to refer to the generation of the new midpoint as the {\it zeroth level} ($l=0$).

One then proceeds to the first level ($l=1$), where the two random positions ${\bf r}^{\prime}_{k+s/4}$ and ${\bf r}^{\prime}_{k+3s/4}$ are sampled from Gaussian distribution functions of semi-width $\sigma_1={\sqrt{s\lambda\delta\tau/4}}$, centered at positions $({\bf r}^{\prime}_{k}+{\bf r}^{\prime}_{k+s/2})/2$ and $({\bf r}^{\prime}_{k+s/2}+{\bf r}^{\prime}_{k+s})/2$. At the next level ($l=2$), one generates 
four new positions and so on. The $l$th level involves the generation of $2^{l}$ new positions, sampled 
from  Gaussian distribution functions of semi-width $\sigma_{l}=\sqrt{2^{-l-1}s\lambda\delta\tau}$.
This ``bisection'' procedure ends when new positions of the particle have been generated at  $k+1,k+2,...,k+s-1$. The last level is obviously $l=m-1$.

The proposed new path $X^{\prime}$ may be either accepted or rejected following an acceptance test based on (\ref{wigg}). It proves much more efficient, however, to break down this global, final acceptance test, into $m-1$ intermediate acceptance tests, each following every level of update. Specifically, after completing the $l$th level one proceeds to the next level with probability
\beq\label{pu}
W(l\to l+1)={\rm exp}\biggl [-\delta\tau \sum_{j\ \epsilon\ level\ l}\biggl (U(R^{\prime}_{k+j})-U(R_{k+j})\biggr )\biggr]\eeq 
aborting the  the process (i.e., rejecting the proposed new path in its entirety) on the first negative outcome of an acceptance test. It is simple to see that the overall acceptance probability for the new path remains the same, given by (\ref{wigg}), on breaking down the acceptance test by levels as explained above. The improvement in efficiency comes from the fact that the final acceptance is mostly influenced by the largest displacements, e.g., that of the midpoint. Thus, 
one can reject early (i.e., after the first level), and with relatively little computational effort, moves that most likely will eventually be rejected anyway. 

The value of $s$ (namely the length of the portion of path that is updated) is set to ensure an optimally efficient sampling. The minimum possible value is $s=2$, which gives the highest acceptance rate, but  at the same time also produces a  modest path update (a single point of the path is modified). This becomes inefficient at low temperature, as paths can be fairly long (e.g., several hundred slices) and such ``single-slice" updating can result in  a very slow diffusion through configuration space, and consequently in undesirably long equilibration and auto-correlation times. It is therefore advantageous to take $s$ as large as possible, while still ensuring a reasonably high acceptance rate for multi-level moves (acceptance rate is a rapidly decreasing function of $s$). In our algorithm, we typically adjust $s$ so that the acceptance rate remains roughly between 20\% and 50\%. 

\subsubsection{``Permute'' type moves}\label{perm}
These moves involve a group of $1 < n \le N$ particles. They are similar to the {\it wiggle} type moves, in that corresponding portions of the paths of the $n$ particles are modified, at $s-1$ consecutive points. An additional aspect, however, is that the modified portion of the path of a particle in the group will connect, at  $k+s$, to the path of a {\it different} particle, among the $n$ selected. This elementary move clearly allows one to sample permutations of the $N$ particles in the system, over the imaginary time interval $[0,\beta]$. 

The basic scheme of the move is as follows: first, a permutation of particle labels at  $k+s$ is sampled; the number $n$ of particles involved in this permutation (henceforth referred to as the {\it cycle}) is not chosen {\it a priori}, but can vary from 2 to $N$. Once the permutation is selected, new single-particle paths are constructed, in a way completely analogous to that used in the ``wiggle'' moves. Finally, the new many-particle path $X^{\prime}$ so obtained is accepted with probability given by (\ref {wig}). Again, it proves convenient to choose a sampling probability for the permutation that will in turn simplify the acceptance probability. Going back to Eq. (\ref{casino}), the first term of the product is now used to sample permutations, whereas the remaining terms are used to construct paths consistent with the permutations that have been sampled.

The sampling of a permutation is a recursive process in which particles are successively added to the cycle. The addition of a single particle includes an acceptance test, and the sampling of the particle from a table. 
One begins by selecting a random particle, say the $\nu$th for definiteness. Based on (\ref{casino}), a table is constructed, $K_{\nu\omega}^{(1)}$, with entries as follows:
\beq\label{tablek}
K^{(1)}_{\nu\omega} = \rho_{F}({\bf r}_{\nu k},{\bf r}_{\omega k+s},s\delta\tau)\ (1-\delta_{\nu\omega})
\eeq
where ${\bf r}_{\nu k}$ is the position of the $\nu$th particle particle at point $k$, whereas ${\bf r}_{{\omega k+s}}$ is that of the $\omega$th particle at point $k+s.$ \cite{note3}
At this point, a first acceptance test is performed, namely the process will continue on to the next stage (i.e., selection of the permutation partner for particle $\nu$) with probability
\beq\label{acce}
C^{(1)} = \frac{Q_{1}}{Q_{1}+\rho_{F}({\bf r}_{\nu k},{\bf r}_{\nu k+s},s\delta\tau)}
\eeq
where $Q_{1}=\sum_{\omega}K^{(1)}_{\nu \omega}$. If the acceptance test fails, then the process is aborted, i.e., the permutation move is rejected. Suppose, instead, that a positive outcome is obtained; an entry $\alpha$ is then sampled from the table $K^{(1)}$, with probability ${\Pi}_{\alpha}=K^{(1)}_{\nu\alpha}/Q_{1}$.  We see from (\ref {tablek}) that particle $\nu$ itself is sampled with probability zero, i.e., the sampling of a ``non-identical'' permutation is forced here.  The particle labeled $\alpha$  is selected as the second member of the permutation cycle being constructed. That means that, in the trial path $X^{\prime}$, the path of particle $\nu$ will go through ${\bf r}_{\nu k}$ at the $k$th point and through ${\bf r}_{\alpha k+s}$ at point $k+s$. At this point, one has to sample a new position of particle $\alpha$ at  $k+s$.  Just as for particle $\nu$, one constructs a table 
\beq
K^{(2)}_{\alpha\omega} = \rho_{F}({\bf r}_{\alpha k},{\bf r}_{\omega k+s},s\delta\tau)\ (1-\delta_{\alpha\omega})
\eeq
and another acceptance test analogous to  (\ref {acce}) is carried out, based on the probability
\beq\label{acce2}
C^{(2)} = \frac{Q_{2}}{Q_{2}+\rho_{F}({\bf r}_{\alpha k},{\bf r}_{\alpha k+s},s\delta\tau)}
\eeq
with $Q_{2}=\sum_{\omega}K^{(2)}_{\alpha\omega}$ (Again, the process is aborted if this acceptance test fails). An entry $\mu$ is sampled with probability ${\Pi}_{\mu}=K^{(2)}_{\alpha\mu}/Q_{2}$; in this case, particle $\nu$ is sampled with finite probability, as the cycle can close on the initial particle, whereas particle $\alpha$ is now excluded from the sampling.
 At this point, if $\mu=\nu$, then the permutation cycle is closed, and it includes two particles, namely $\nu$ and $\alpha$. If, on the other hand,
$\mu\ne \nu$, then one must find another particle $\gamma$, which will become a member of the cycle, such that ${\bf r}_{\mu k}={\bf r}_{\gamma k+s}$. 
Again, one constructs a table $K^{(3)}$ as above, the only difference being that now both $\alpha$ {\it and} $\mu$ are excluded from consideration, as both ${\bf r}_{\alpha k+s}$ and ${\bf r}_{\mu k+s}$ are already taken:
\beq
K^{(3)}_{\mu\omega} = \rho_{F}({\bf r}_{\mu k},{\bf r}_{\omega k+s},s\delta\tau)\ (1-\delta_{\alpha\omega}-\delta_{{\mu\omega}})
\eeq
A new acceptance/rejection test is performed, based on a probability $C^{(3)}$ defined analogously to (\ref{acce})-(\ref{acce2}):
\beq
C^{(3)} = \frac{Q_{3}}{Q_{3}+\rho_{F}({\bf r}_{\mu k},{\bf r}_{\mu k+s},s\delta\tau)
+\rho_{F}({\bf r}_{\mu k},{\bf r}_{\alpha k+s},s\delta\tau) }
\eeq
with $Q_{3}=\sum_{\omega}K^{(3)}_{\mu\omega}$,
and in case of success, one proceeds to sample entry $\gamma$ from table $K^{(3)}$,  with probability ${\Pi}_{\gamma}=K^{(3)}_{\mu\gamma}/Q_{3}$.

The basic idea should now be clear: This procedure is iterated until the cycle is finally closed, namely until particle $\nu$ is obtained from the sampling of the table $K^{(n-1)}$. 
Two fundamental aspects of the above scheme to sample permutation cycles, are the exclusion from the tables $K^{(n)}$ of entries corresponding to particles  already in the cycle (the $\nu$th is only excluded from $K^{(1)}$), and the acceptance tests based on $C^{(n)}$, preceding each new particle selection. The sum at the denominator of $C^{(n)}$ includes, besides $Q_n$, free-particle density matrices associated to all the entries excluded in the sampling table $K^{(n)}$.

Once a complete cycle has been obtained,  one must construct trial paths for all particles in the cycle, consistent with the selected new positions at slice $k+s$. This second part is done in exactly the same way as for the ``wiggle'' moves, using the same sampling probabilities. Specifically, new midpoint positions are first sampled for all particles in the cycles;  
then, new positions at  $k+s/2$ and $k+3s/4$ are sampled, and so on, with acceptance tests as in (\ref {pu}) after each level of path update. Note that the values of $U$ at points $k$ and $k+s$ remain unchanged, as no particle is displaced at these points; only particle labels are altered at point $k+s$. It is a simple matter to show that the path sampling probability arising from the above scheme is indeed consistent with (\ref {basic}).

The above scheme to sample permutations is similar to one described in Ref. \onlinecite{ceperley95}; the main difference, possibly significant, is that, in our procedure,  one need {\it not} include, in any of the acceptance tests, a sum of terms representing  all $n$ starting points of the cyclic permutation. Moreover, in our method $n$ distinct particles are sampled by construction, and the ``identity'' permutation, namely the one in which particle labels are left unchanged from $k$ to $k+s$, is excluded from the sampling. 

Just as in the ``wiggle'' moves, $s$ must be chosen appropriately, namely, long enough that non-trivial permutation cycles can be sampled with appreciable probability. If $s$ is small, particularly if one is working with a small value of $\delta\tau$, the functions $\rho_F$ are negligible small for distances of the order of the average distance between particles, rendering it exceedingly unlikely to  go beyond the first acceptance test (Eq. \ref{acce}).
On the other hand, taking $s$ too long, while allowing  for large permutation cycles being sampled,  results in very low overall acceptance for these cycles, much for the same reasons why acceptance falls for the ``wiggle'' moves as well if $s$ is too large. In the calculations whose results are illustrated in the next section, we have generally found that  the optimal choice of $s$ is generally the same as for the ``wiggle'' moves.

Even when $s$ is optimally chosen,  typical values of acceptance for permutations are low.  One tries to keep the efficiency reasonable by attempting a large number  permutational moves, which can be done fairly rapidly within the relatively simple scheme outlined above. We typically attempt several tens of thousands of permutations  between two consecutive sets of ``wiggle'' type moves, in which full updates of the paths of all particles are attempted.

\section{Results}
We now describe the results of our PIMC simulations of condensed bulk \he4 at low temperature (1 K $\le T\le $ 4K), obtained with the algorithm described in the previous section. The model Hamiltonian for the system of interest is given by (\ref {ham}). For the purpose of comparing our results with existing calculations,\cite{ceperley95,ceperley86} we used an early version of the Aziz potential\cite{aziz} to describe the interaction between a pair of \he4 atoms ($\lambda$=6.0596 K\AA$^{2}$). 

We have computed several energetic and structural properties of the system. We have observed convergence of the energy estimates  with a value of  the ``imaginary time step''  $\delta\tau$=1/640 K$^{-1}$.  All of the results presented in this section are obtained with this value of $\delta\tau$. We estimate any residual, systematic error on the energy arising from our  path discretization, to be worth no more than 0.1 K per \he4 atom. For structural properties, we have observed that estimates obtained with (up to four times) larger values of $\delta\tau$ are indistinguishable, within statistical uncertainties, from those obtained with the above-mentioned value of $\delta\tau$. 

We found our optimal value of $s$, for both ``wiggle'' and ``permute'' type moves, to be $s=6$.\cite{notec}  Accordingly, the length of the portion of path that is updated on each move is $2^{s}=64$ imaginary time slices, which corresponds to an imaginary time interval of 0.1 K$^{-1}$, with our choice of time step. 
Unless otherwise stated, the number of particles in our simulated system is $N$=64, as in Ref. \onlinecite{ceperley86}, but results for $N$=216 were obtained as well..  
\subsection{Energetics}
The energy estimators utilized in this work are described, for instance, in Ref. \onlinecite{voth}. Specifically, the average kinetic energy per particle $K$ is obtained as 
\begin{eqnarray}\label{kin}
\langle K \rangle \approx \frac{1}{2\delta\tau}-\frac{1}{4\lambda\delta\tau^{2}}\biggl \langle \biggl ( {\bf r}_{k}-{\bf r}_{k+1}\biggr )^{2} \biggr \rangle+\frac{\lambda\delta\tau^2}{9}
\biggl \langle \biggl ( \nabla V(R_{2k}\biggr )^{2}\biggr \rangle
\end{eqnarray}
where $\langle ...\rangle$ stands for statistical average, $({\bf r}_{k}-{\bf r}_{k+1})^{2}$ is the square 
distance between the positions of a particle at adjacent points along the path, whereas the gradient of the potential energy in the third term is taken with respect to the coordinate of one of the particles, at an ``even'' slice.  The potential energy per particle $\langle v\rangle$ is instead obtained as
\beq\label{potl}
\langle v \rangle \approx {1\over N}\biggl \langle V(R_{2k-1})\biggr \rangle
\eeq

Both relations (\ref{kin}) and (\ref{potl}) are approximate, approaching the exact results only in the limit $L\to\infty$, $\delta\tau\to 0$. The above kinetic energy estimator is not the most efficient; it is known that the so-called ``virial'' estimator yields more accurate results (i.e., smaller statistical errors), given the same amount of computing time.\cite{cha99} However, the estimator (\ref{kin}) has been most commonly adopted in previous calculations of this type.
 In all of our calculations, we estimated the contribution to the potential energy attributed to particles outside the main simulation cell by assuming that the pair correlation function $g(r)$ equals one outside the cell.
 
Table \ref{tab1} summarizes our results for the energetics of bulk \he4, at different temperatures. Shown in parentheses are the corresponding results from Ref. \onlinecite{ceperley86}. The estimates are in quantitative agreement,  taking into account  the statistical uncertainties of the two calculations. 
Amusingly, our statistical errors  on the kinetic energy are not much smaller than those of Ref. \onlinecite{ceperley86}, in spite of the fact that our calculation benefits of two more decades of advances in computing hardware. This is not too surprising, however, as the calculation of the kinetic energy, especially based on the estimator (\ref{kin}), is known\cite{cha99} to be the place where the limitations are most evident of using less than optimal a high-temperature density matrix $\rho_\circ$, such as the one used in this work. Still, the results of our calculation seem altogether satisfactory, giving us confidence that our PIMC implementation, including the new permutation sampling engine, performs correctly.
\subsection{Single-particle diffusion in imaginary time}
We observe excellent agreement between our results and those of Ref. \onlinecite{ceperley86} for structural properties, such as the pair correlation function (an example is given in Fig. 1); however, effects of quantum statistics on these quantities are small,\cite{ceperley95} and therefore their computation does not provide a particularly significant test of an algorithm to simulate indistinguishable quantum particles.

More telling are measures of the diffusion of particles in imaginary time. Fig. 2 shows results for the quantity $D(\tau)$, defined as
\beq\label{dif}
D(\tau)=\frac{\biggl \langle \biggl ( {\bf r}(\tau)-{\bf r}(0)\biggr )^{2}\biggr \rangle}{6\lambda\tau}
\eeq
where $\langle ...\rangle$ stands for statistical average.  The two curves shown in the figures represent values of $D$ computed by PIMC for bulk \he4 (solid line), as well as for a system of {\it distinguishable} \he4 atoms, both at a temperature $T$=2 K. While in the first case \he4 atoms are treated as bosons, and therefore permutations are included, in the second case  no permutations of particles are allowed. 

Obviously, because in the latter case one must have ${\bf r}(\beta)={\bf r}(0)$, i.e., single-particle paths must close onto themselves, it must be $D(\beta)=0$. On the other hand, if permutations are allowed, then single-particle paths can become entangled, and  $D(\beta)$ may take on a finite value. Moreover, the value of $D$ is greater, at all imaginary times, in the case of Bose statistics; this is fairly intuitive, as the fact that particles are indistinguishable enhances the degree of delocalization of each individual particle.
\subsection{Superfluid density}
We have also computed the \he4 superfluid density $\rho_S$, using the well-known ``winding number'' estimator.\cite{pollock87}  At the lowest temperature considered in this work, namely $T$=1.1765 K, our result is $1.02\pm0.10$, which is in agreement with experiment and with the PIMC result of Ref. \onlinecite{pollock87}. We obtained this result with a number of slices $L=544$. It should also be mentioned that it appears possible to obtain a reasonably accurate estimate of $\rho_{S}$ using considerably fewer imaginary  time slices (the result obtained with $L=136$ is indistinguishable, within statistical uncertainties, from the one quoted above), and that reducing $L$ also causes a significant reduction of the statistical error on $\rho_{S}$.  In general, however, if $L$ becomes relatively large, namely of the order of a few hundred, lengthy simulations are required in order to reduce statistical error  to an acceptable size (e.g.,  0.05 or less). This problem seems common to other PIMC implementations as well,  and it is not clear to us to what extent it may signal an inefficiency of our sampling method.
\subsection{Statistics of Permutations and Permutation Cycles}
In order to characterize the performance of the permutation sampling algorithm, one may also look at quantities easily accessible in a simulation, which may not directly relate to anything measurable but provide a possible baseline for comparison of different algorithms. Table 2 provides statistics of permutation acceptance for a PIMC simulation of 64 \he4 atoms at $T$=1.1765 K (the lowest temperature considered here). The total number of permutations attempted in this run is $4.5\times 10^{5}$, and the fraction of accepted permutations (of any cycle length) is approximately 0.4\%. Permutations were sampled over an imaginary time interval of length 0.1 K$^{-1}$. \\

As one can see from the second column, 2-particle permutations are sampled overwhelmingly more than others; however, the rate of acceptance of attempted permutations is essentially constant, independent of $n$. This is found to be the case at all temperatures considered in this study. One may think that it would be advantageous to increase the rate at which permutations of more than two particles are sampled, since they presumably enhance the diffusion of the random walk through path space. Indeed, it is straightforward to generalize our sampling algorithm, so that  permutations including more than two particles will be sampled more often. In practice, however, we found that including pair permutations is beneficial, in that it leads to a greater overall rate of acceptance of attempted permutations. How much of this is problem- or algorithm-dependent is difficult to say. 
Both quantities shown in Table 2 are rapidly decreasing functions of the temperature, as expected. \\
Although they are not sampled directly, permutation cycles involving large numbers of particles can and do occur, as a result of sampling many permutations involving few particles.  Fig. 3 shows a histogram of probability for a particle to be part of a permutation cycle of length $n$ (i.e., involving $n$ particles) in a PIMC simulation carried out with the methodology illustrated above, at three different low temperatures below the $\lambda$-transition. Although the data are somewhat noisy, these results are in quantitative agreement with those of Ceperley \cite{ceperley95} for the same system, using the CP methodology. As the temperature is lowered, the probability that a particle will belong to a cycle of length $n$ becomes independent of $n$.

\section{Conclusions}
A new algorithm to perform the sampling of permutations of indistinguishable particle in Path Integral Monte Carlo simulations was introduced. This procedure is similar, in spirit, to existing methods, but differs in some important aspects, and may have some advantages. We have tested it by performing a PIMC simulation of liquid \he4 at low temperature, in the superfuid regime. Aside from the permutation sampling scheme, the rest of the  PIMC methodology utilized here is {\it not} optimized for \he4 calculations. In particular, it is worth repeating that much better options exist for the high-temperature density matrix, which can drastically reduce the number of time slices needed for convergence. Still, the calculation proves quite feasible with currently available, moderately powerful workstations. It should also be noted that, while the use of a more accurate high-temperature density matrix (specifically, the pair-product approximation (\ref{ppa})) greatly enhances the efficiency of calculations for a highly quantal, hard-sphere-like system such as helium, for other condensed systems such as molecular hydrogen, which feature a lesser degree of zero-point motion, or Coulomb system, for which the interaction potential is considerably less ``stiff'', the high-temperature density matrix utilized here and Eq. (\ref{ppa})  may be of comparable efficiency.\\
For comparison with existing calculations we limited the size of the system studied to $N$=64 particles, but it should be mentioned that simulations of systems with as many as four times more particles are also possible, with a reasonable amount of computer time (of the order of a month per thermodynamic point). \\
We have attempted to furnish as much quantitative information as possible, that may help assess the relative efficiency of the permutation scheme proposed here against existing ones. Obviously, a direct comparison of results provided by implementations only differing by the permutation scheme adopted, is also desirable. It is our hope that such a comparison will be soon carried out.
\section*{Acknowledgments}
This work was supported in part by the Petroleum Research Fund of the American Chemical Society under research grant 36658-AC5, by the Natural Sciences and Engineering Research council of Canada (NSERC) under research grant G121210893. The author wishes to acknowledge useful discussions with Saverio Moroni.

\newpage
\section*{TABLES}
 \begin{table}[ht]
 \caption{Kinetic and potential energy per atom  of condensed bulk  \he4, computed by PIMC at different thermodynamic conditions. Third column shows the number of imaginary time slices utilized in the calculation. The simulated system comprises 64 particles. Results in parentheses are 
from Ref. 2. }
{\begin{tabular}{@{}ccccc@{}} \toprule 
$T$ (K) &$\rho$ (\AA$^{-3}$) &$L$ & kinetic (K) & potential (K)\\ \colrule
1.1765  &0.02182 & 544 &$14.123\pm0.028$ & $-21.3127\pm0.0025$  \\
      & &                                        &($14.17\pm0.08$) &($-21.35\pm0.04$)  \\
      1.379 &0.02182 &466 &$14.201\pm0.024$ &$-21.3195\pm0.0029$  \\
      & &                                      &$(14.23\pm0.08)$ &$(-21.35\pm0.08)$  \\
 1.600 &0.02183 &400 &$14.334\pm0.034$ & $-21.3435\pm0.0037$  \\
 & & &($14.40\pm0.08$)&($-21.39\pm0.04$)  \\
 1.818 &0.02186 &352 &$14.468\pm0.059$ &$-21.3913\pm0.0060$ \\
 & & &($14.71\pm0.08$)&($-21.44\pm0.04$) \\
 2.000 &0.02191 &320 &$14.862\pm0.071$&$-21.4810\pm0.0078$ \\
 & & &($15.05\pm0.08$)&$(-21.57\pm0.04)$ \\
 2.353 &0.02191 &272 &$15.821\pm0.062$&$-21.5791\pm0.0072$ \\
 & & &($15.75\pm0.08$) &($-21.60\pm0.04$)  \\ \botrule
\end{tabular}\label{tab1}}
\end{table}
\newpage
\begin{table}[h]
\caption{Permutation statistics for a PIMC simulation of bulk liquid \he4 at T=1.1765 K and a density $\rho=0.02182$ \AA$^{-3}$. The second column indicates the percentage of {\it attempted} permutations involving $n$ particles, sampled as explained in the text. The third column yields the percentage of attempted permutations that are {\it accepted}. The total number of attempted permutations for this run is $4.5\times 10^{5}$.}
{\begin{tabular}{@{}c|c|c@{}} \toprule 
$n$  & \% attempted & \% accepted  \\ 
1 & 0.0 & 0.0  \\
2 & 82.2 &0.38  \\
3 &12.5 &0.46  \\ 
4 & 3.5 & 0.27 \\
5 &0.1 & 0.54 \\
$> 5$ & 1.7 & 0.00 \\
\botrule
\end{tabular}\label{tab2}}
\end{table}
\newpage
\section*{FIGURE CAPTIONS}
\begin{figure}[h]
\caption{Pair correlation function $g(r)$ computed by PIMC for liquid $^4$He at a density $\rho$=0.02182 \AA$^{-3}$ and at a temperature $T$=1.1765 K.  The simulated system comprises 216 particles. Statistical errors are smaller than the size of the symbols. The effects of quantum statistics are barely visible on the scale of the figure.}
\end{figure}
\begin{figure}[h]
\caption{Imaginary time diffusion coefficient $D(\tau)$, defined in Eq. (\ref{dif}), computed by PIMC for bulk \he4 (solid line) and for an ensemble of distinguishable (i.e., no permutations allowed) \he4 atoms at the same density (dashed line). The temperature is $T$=2 K. Statistical errors on both curves are of the order of $10^{-3}$ or less. The number of particles in the system is $N$=64.}
\end{figure}
\begin{figure}[h]
\caption{Probability for a single particle to belong to a permutation cycle including $n$ particles, in a PIMC simulation of 64 \he4 atoms at saturated vapor pressure and at three different low temperatures.}
\end{figure}
\newpage\begin{figure}[h]
\centerline{\includegraphics[height=6.5in,angle=-90]{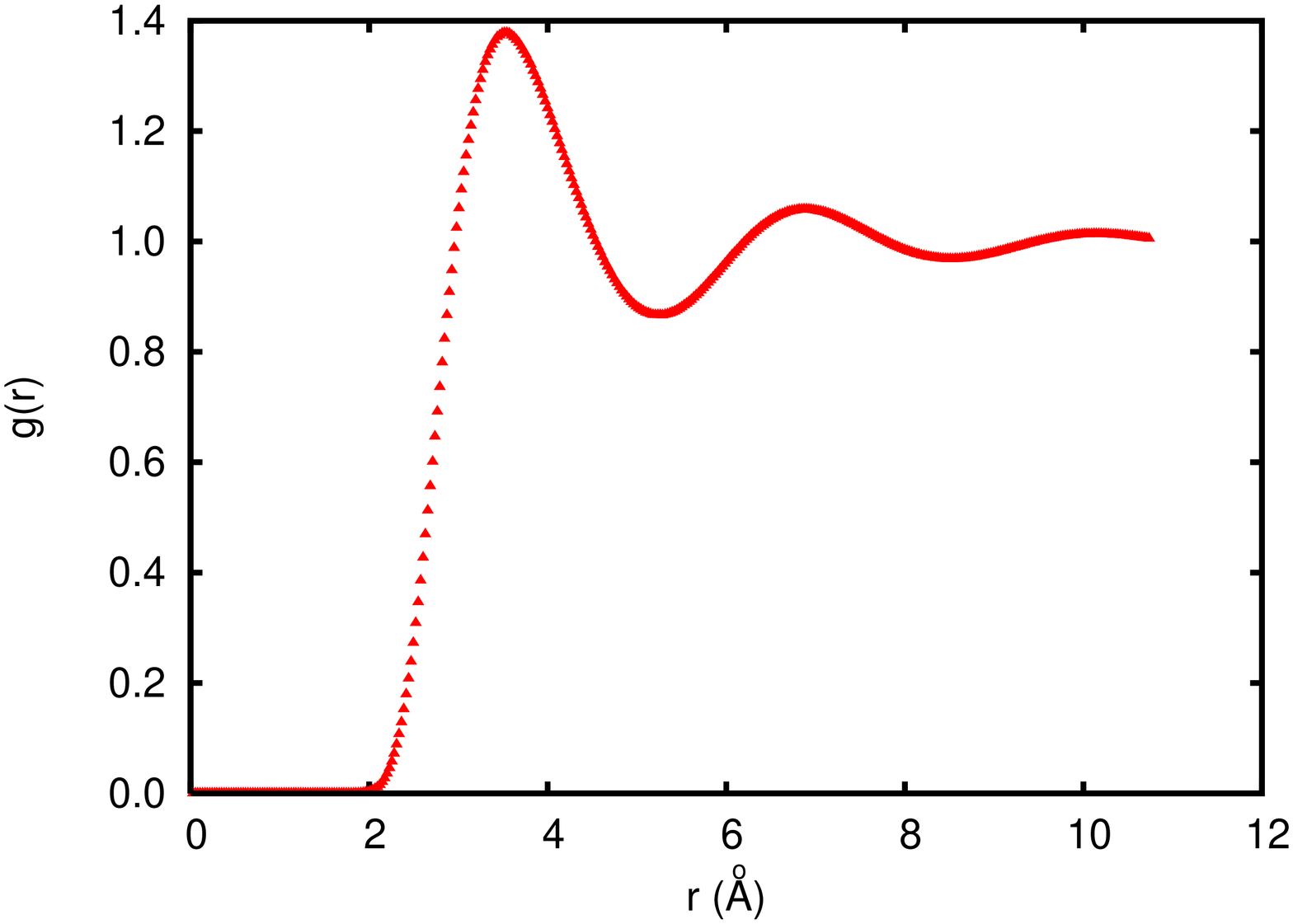}}
\small{Fig. 1}
\end{figure}
\begin{figure}[h]
\centerline{\includegraphics[height=6.5in,angle=-90]{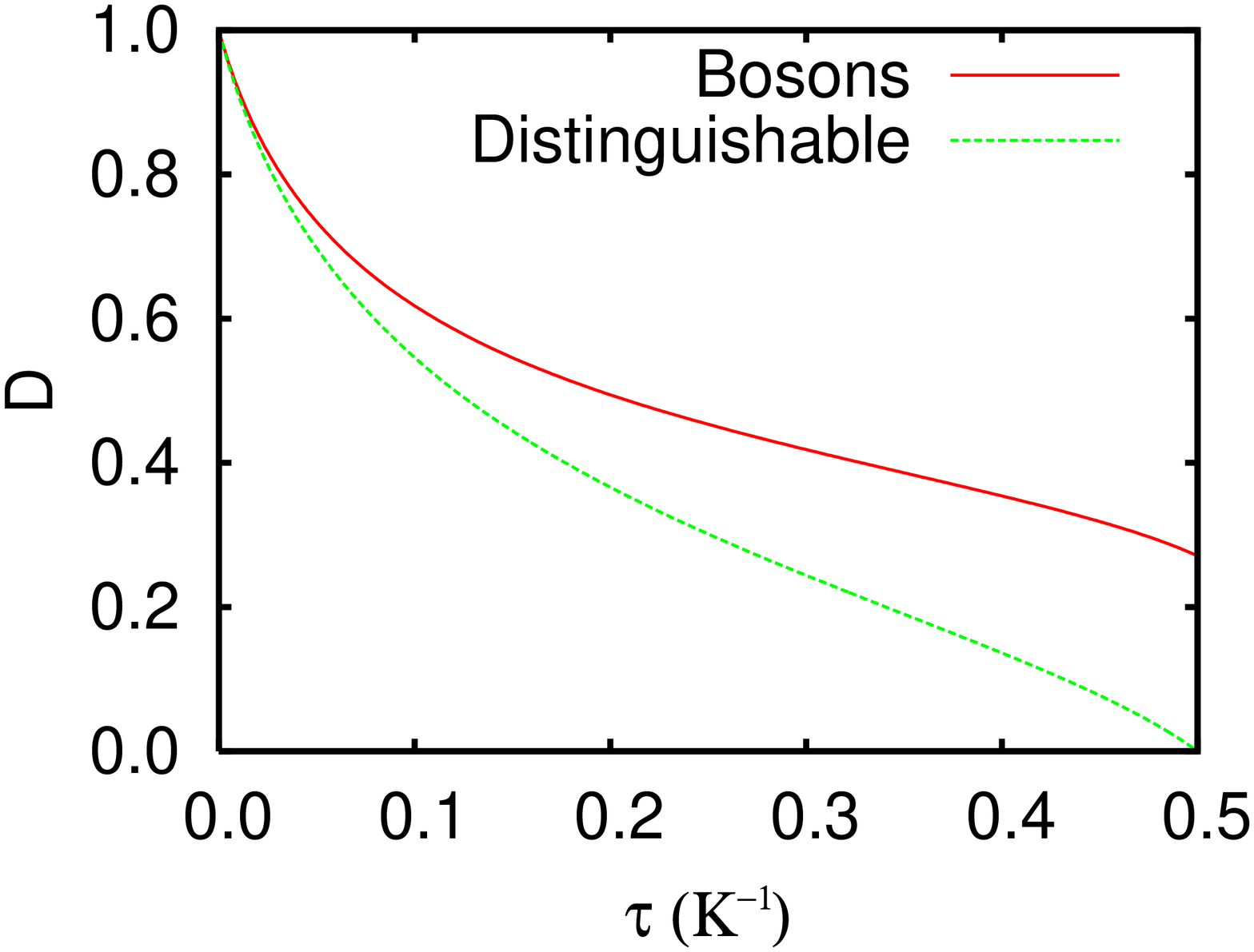}}
\small{Fig. 2}
\end{figure}
\begin{figure}[h]
\centerline{\includegraphics[height=6.5in,angle=-90]{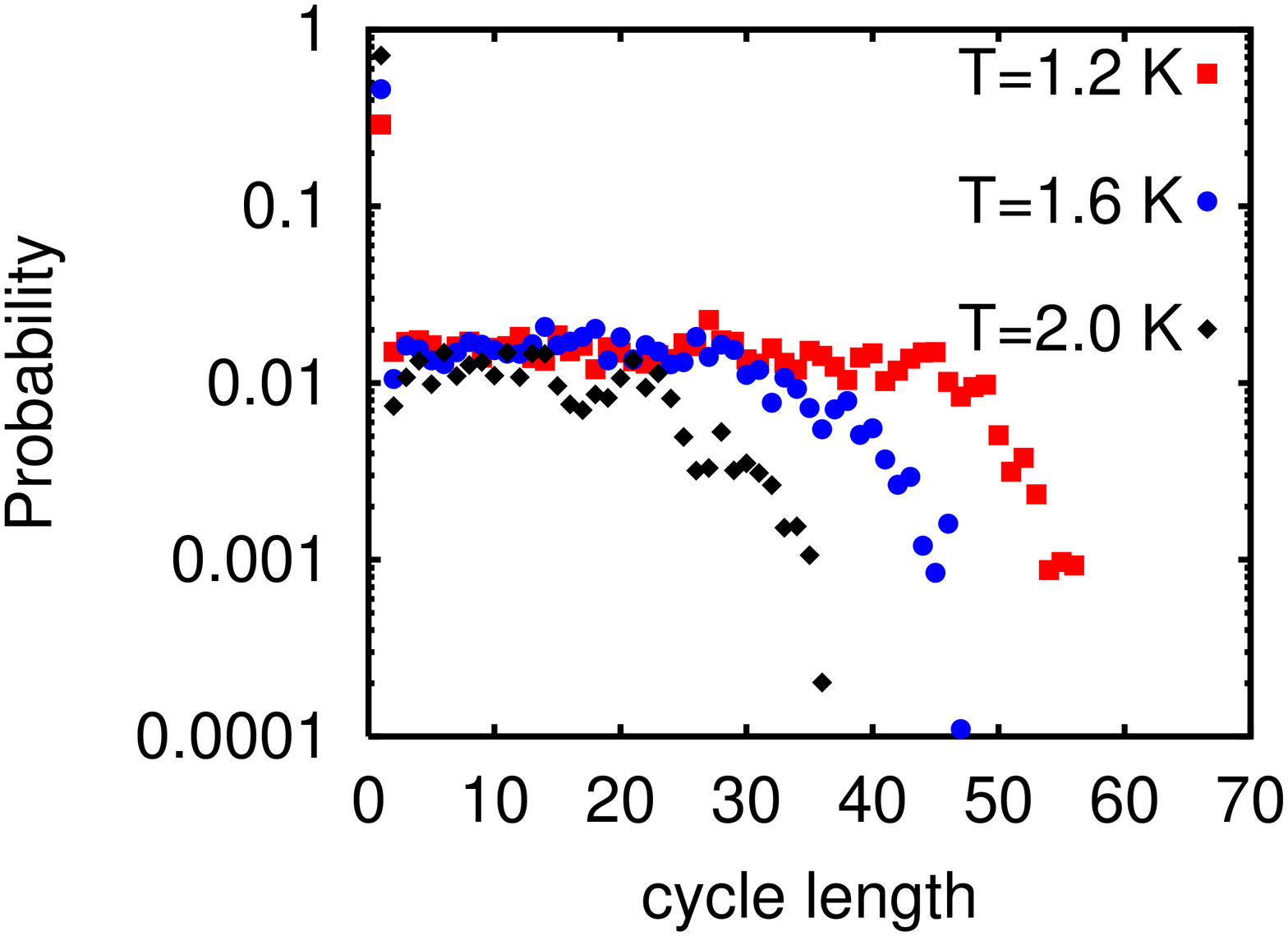}}
\small{Fig. 3}
\end{figure}

\begin{thebibliography}{00}
\bibitem{note2}
It all can be derived using the identity $(a-b)^2+(b-c)^{2}=(a-c)^{2}/2+2(b-\bar b)^{2}$, where
$\bar b=(a+c)/2$.
\bibitem{note3}
We emphasize that all distances in (\ref{tablek}) are computed by taking the periodic image of each particle that is closest to particle $\nu$. Thus, as a result of permutation sampling, paths cross over into adjacent simulation cells. This aspect is crucial in estimating the superfluid density, or imaginary time diffusion of individual particles. For details, see Ref. \onlinecite{ceperley95}.
\bibitem{notec}
For ``permute'' type moves, we define the optimal value of $s$ that for which the rate of acceptance of attempted permutations is greatest.
\end{thebibliography}
\end{document}